\input epsf
\documentclass[12pt]{article}
\begin{document}

\def\bd{\begin{displaymath}}\def\ed{\end{displaymath}}
\def\be{\begin{equation}}\def\ee{\end{equation}}
\def\bea{\begin{eqnarray}}\def\eea{\end{eqnarray}}
\def\nn{\nonumber}\def\lb{\label}

\def\a{\alpha}\def\b{\beta}\def\c{\chi}\def\d{\delta}\def\e{\epsilon}
\def\f{\phi}\def\g{\gamma}\def\h{\theta}\def\i{\iota}\def\j{\vartheta}
\def\k{\kappa}\def\l{\lambda}\def\m{\mu}\def\n{\nu}\def\o{\omega}\def
\p{\pi}\def\q{\psi}\def\r{\rho}\def\s{\sigma}\def\t{\tau}\def\u{\upsilon}
\def\vu{\varphi}\def\w{\varpi}\def\y{\eta}\def\x{\xi}\def\z{\zeta}

\def\D{\Delta}\def\F{\Phi}\def\G{\Gamma}\def\H{\Theta}\def\L{\Lambda}
\def\O{\Omega}\def\P{\Pi}\def\Q{\Psi}\def\S{\Sigma}\def\U{\Upsilon}\def\X{\Xi}

\def\lie{{\cal L}}\def\de{\partial}\def\na{\nabla}\def\per{\times}
\def\inf{\infty}\def\id{\equiv}\def\mo{{-1}}\def\ha{{1\over 2}}
\def\qu{{1\over 4}}\def\pro{\propto}\def\app{\approx}
\def\we{\wedge}\def\di{{\rm d}}\def\Di{{\rm D}}

\def\Ei{{\rm Ei}}\def\li{{\rm li}}\def\const{{\rm const}}\def\ex{{\rm e}}
\def\arcsh{{\rm arcsinh}}\def\arcch{{\rm arccosh}}
\def\arcth{{\rm arctanh}}\def\arccth{{\rm arccoth}}
\def\diag{{\rm diag}}

\def\gmn{g_{\m\n}}\def\ep{\e_{\m\n}}\def\ghmn{\hat g_{\m\n}}\def\mn{{\mu\nu}}
\def\dix{\int d^2x\ \sqrt{-g}\ }\def\ds{ds^2=}\def\sg{\sqrt{-g}}
\def\dhx{\int d^2x\ \sqrt{-\hat g}\ }\def\dex{\int d^2x\ e\ }

\def\tors#1#2#3{T_{#1#2#3}}\def\curv#1#2#3#4{R_{#1#2#3#4}}
\def\af{asymptotically flat }\def\hd{higher derivative }\def\st{spacetime }
\def\fe{field equations }\def\bh{black hole }\def\as{asymptotically }
\def\eqs{equations }\def\eom{equations of motion }
\def\tran{transformation }\def\ther{thermodynamical }\def\coo{coordinates }
\def\bg{background }\def\gs{ground state }\def\bhs{black holes }
\def\sc{semiclassical }\def\hr{Hawking radiation }\def\sing{singularity }
\def\ct{conformal transformation }\def\cc{coupling constant }
\def\crel{commutation relations }\def\tl{transformation law }
\def\ns{naked singularity }\def\gi{gravitational instanton }
\def\rep{representation }\def\gt{gauge transformation }
\def\cco{cosmological constant }\def\em{electromagnetic }
\def\ssy{spherically symmetric }\def\cf{conformally flat }
\def\cur{curvature }\def\tor{torsion }\def\ms{maximally symmetric }
\def\coot{coordinate transformation }\def\diff{diffeomorphisms }
\def\gct{general coordinate transformations }\def\gts{gauge transformations }
\def\pb{Poisson brackets }\def\db{Dirac brackets }\def\ham{Hamiltonian }
\def\cd{covariant derivative }\def\dof{degrees of freedom }
\def\hdim{higher dimensional }\def\ldim{lower dimensional }
\def\SR{special relativity }
\def\dys{dynamical system }\def\cps{critical points }\def\dim{dimensional }
\def\sch{Schwarzschild }\def\min{Minkowski }\def\ads{anti-de Sitter }
\def\RN{Reissner-Nordstr\"om }\def\RC{Riemann-Cartan }\def\poi{Poincar\'e }
\def\KK{Kaluza-Klein }\def\pds{pro-de Sitter }\def\des{de Sitter }
\def\BR{Bertotti-Robinson }\def\MP{Majumdar-Papapetrou }
\def\GR{general relativity }\def\GB{Gauss-Bonnet }\def\CS{Chern-Simons }
\def\EH{Einstein-Hilbert }\def\EPG{extended \poi group }
\def\JT{Jackiw-Teitelboim }
\def\dpa{deformed \poi algebra }\def\psm{Poisson sigma model }
\def\td{two-dimensional }\def\trd{three-dimensional }
\def\lt{Lorentz transformations }\def\com {center of mass }

\def\PL#1{Phys.\ Lett.\ {\bf#1}}\def\CMP#1{Commun.\ Math.\ Phys.\ {\bf#1}}
\def\PRL#1{Phys.\ Rev.\ Lett.\ {\bf#1}}\def\AP#1#2{Ann.\ Phys.\ (#1) {\bf#2}}
\def\PR#1{Phys.\ Rev.\ {\bf#1}}\def\CQG#1{Class.\ Quantum Grav.\ {\bf#1}}
\def\NP#1{Nucl.\ Phys.\ {\bf#1}}\def\GRG#1{Gen.\ Relativ.\ Grav.\ {\bf#1}}
\def\JMP#1{J.\ Math.\ Phys.\ {\bf#1}}\def\PTP#1{Prog.\ Theor.\ Phys.\ {\bf#1}}
\def\PRS#1{Proc.\ R. Soc.\ Lond.\ {\bf#1}}\def\NC#1{Nuovo Cimento {\bf#1}}
\def\JP#1{J.\ Phys.\ {\bf#1}} \def\IJMP#1{Int.\ J. Mod.\ Phys.\ {\bf #1}}
\def\MPL#1{Mod.\ Phys.\ Lett.\ {\bf #1}} \def\EL#1{Europhys.\ Lett.\ {\bf #1}}
\def\AIHP#1{Ann.\ Inst.\ H. Poincar\'e {\bf#1}}\def\PRep#1{Phys.\ Rep.\ {\bf#1}}
\def\AoP#1{Ann.\ Phys.\ {\bf#1}}
\def\grq#1{{\tt gr-qc/#1}}\def\hep#1{{\tt hep-th/#1}}

\def\mm{m_1+m_2}\def\xx{x_1-x_2}\def\kk{{2\k^2}}\def\km{{\k^2m}}
\def\wr{with respect to }

\begin{titlepage}
\vspace{.3cm}
\begin{center}
\renewcommand{\thefootnote}{\fnsymbol{footnote}}
{\Large \bf Two-dimensional static black holes with pointlike sources}
\vskip 15mm
{\large \bf {M.~Melis\footnote{email: mmelis@unica.it}}} and
{\large \bf {S.~Mignemi\footnote{email: smignemi@unica.it}}}\\
\renewcommand{\thefootnote}{\arabic{footnote}}
\setcounter{footnote}{0}
\vskip 7mm
{\small
Dipartimento di Matematica, Universit\`a di Cagliari,\\
Viale Merello 92, 09123 Cagliari, Italy\\
\vspace*{0.4cm}
 INFN, Sezione di Cagliari\\
}
\end{center}
\vfill
\centerline{\bf Abstract}
\vskip 15mm

We study the static \bh solutions of generalized two-dimensional
dilaton-gravity theories generated by pointlike mass sources,
in the hypothesis that the matter is conformally coupled.

We also discuss the motion of test particles. Due to conformal coupling,
these follow the geodesics of a metric obtained by rescaling the
canonical metric with the dilaton.

\vfill
\end{titlepage}

\section{Introduction}
Two-dimensional gravity models have been widely investigated in
last years as toy models for higher dimensions world \cite{KL}.
However, they differ in an important respect from higher-dimensional
models, because of the necessity of introducing a scalar field $\y$
(dilaton) in order to obtain nontrivial field equations \cite{JT}.
Although some authors consider it
simply as an auxiliary field, the dilaton plays a role in the
interpretation of the theory. For example, the zeroes of
the dilaton can be interpreted as singularities of the geometry
\cite{CM}. Moreover, the dilaton field admits arbitrary kinetic and
potential terms, and this opens the possibility for the existence
of many inequivalent models of two-dimensional gravity.

Of special interest is of course the study of \bh solutions. These have
been
obtained for a variety of models, but usually without reference to the
specific matter sources. In this paper, we study the solutions generated
by static pointlike particles for the models with power-law dilaton
potential introduced in ref.\ \cite{Mi}.
The same problem was previously considered in ref.\ \cite{MS} in the
particular case of a linear dilaton potential, but the role of the
dilaton was disregarded in that paper. Although the equations for
the metric can be solved independently in that special case, the
dilaton equations add some constraints on the parameters of the
solution. This introduces some difference between our results and
those of ref.\ \cite{MS}. Also the thermodynamics is modified if the
contribution of the dilaton is taken into account properly.

We also consider the motion of test particles.
As we shall see, for conformal coupling of the sources, the
newtonian potential grows linearly with the distance, and hence
the gravitational force is constant. However,
if one assumes for consistency that the test particles are also
conformally coupled, they will follow the geodesics of a rescaled
metric, and hence the force experienced is modified.

\section{Single source}
We consider the model of ref.\ \cite{Mi} with conformally coupled matter.
Conformal coupling appears to be the most natural in this context,
since it gives rise to gravitational \fe which relate the geometry to
the matter in a fashion similar to their higher-dimensional counterparts.

We start from the action
\be\lb{act}
I={1\over\kk}\int d^2x\sqrt{-g}\left(\y\,R+\l^2\y^h+\kk\,\y\,L_M\right),
\ee
with $h$ a positive integer, $\l^2$ a "cosmological" constant, and
$2\k^2$ the (dimensionless) gravitational constant.
Since we are interested in \bh
solutions, we only consider the case $\l^2>0$, for which the
metric function $g(x)$ defined below is positive asymptotically.
Varying the action (\ref{act}) yields the \fe
\bea
&&R+\l^2h\y^{h-1}=-\kk\,L_{M},\lb{e1}\\
&&-(\na_\mu\na_\nu-g_\mn\na^2)\eta-{\l^2\over2}\, g_\mn\,\eta^h
=\k^2\,\y\,T_\mn,\lb{e2}
\eea
where, for a free particle of mass $m$ located at $x_0$,
\be\lb{LM}
L_M=-m\d(x-x_0),\qquad T_\mn=m\d(x-x_0)u_\m u_\n,
\ee
with $u^\m=dx^\m/ds$ and $u^\m u_\m=-1$ \cite{Di}.
Note that in two dimensions the sign of $m$ is not fixed a priori.

We adopt static \coo, such that
\be
ds^2=-g(x)dt^2+g^\mo(x)dx^2,\qquad \y=\y(x),
\ee
and consider a particle at rest at the origin, for which
$T_{00}=m\,g^\mo\d(x)$, $T_{11}=0$.
The \fe can then be put in the form
\bea
\lb{fe1}&&g''-\l^2h\y^{h-1}=-\kk m\,\d(x),\\
\lb{fe2}&&g\y''=-\k^2 m\,\d(x)\,\y,\\
\lb{fe3}&&g'\y'=\l^2\y^h,
\eea
where a prime denotes a derivative \wr $x$.

When $m=0$, the solutions are well known \cite{Mi}.
Of course, the solutions maintain the same form when $m\ne0$, except
at the location of the point masses, where the derivative of the
metric has a discontinuity proportional to the mass of the source.
It is easy to see that the integration
of (\ref{fe1}) yields a potential for the point particle linear in
the distance. This corresponds to a constant gravitational force
for minimally coupled test particles.

We start our investigation from the special cases $h=0$ and $h=1$.

\subsection{$h=0$}
In this case spacetime is everywhere flat, except at $x=0$.
We make the ansatz
\be
\y=\a|x|+\y_0,\qquad g=\b|x|+\g.
\ee
Substituting in (\ref{fe1})-(\ref{fe3}), we obtain the conditions
\bd
2\b=-2\km,\qquad 2\a\g=-\km\y_0\qquad\a\b=\l^2,
\ed
which are easily solved, giving
\bd
\b=-\km,\quad\a=-{\l^2\over\km},\quad\g={(\km)^2\over2\l^2}\,\y_0,
\ed
and hence
\be\lb{h0}
\y=-{\l^2\over\km}|x|+\y_0,\qquad
g=-\km|x|+{(\km)^2\over2\l^2}\,\y_0.
\ee
Regular black hole solutions exist therefore only for $m<0$ and
$\y_0<0$.
The horizons of the \bh are located at $x_0=\pm\km\,\y_0/2\l^2$.

One can calculate the ADM mass of the \bh by means of the Mann formula
\cite{Ma}, which for the models (1) reads\footnote{In the formula
reported in \cite{Mi} a factor of $h+1$ was missing.}
\bd
M={1\over2\a}\left({\l^2\over h+1}\,\y^{h+1}-g\,\y'^2\right).
\ed
Substituting the solution (\ref{h0}), one obtains $M=-{1\over4}\,\km
\y_0$.
The temperature of the horizon can be obtained in the standard way as
$T=|g'(x_0)|/4\p$ and reads $T={\k^2|m|\over 4\p}\propto M$.
Contrary to the \sch black hole, it vanishes for $M\to0$.
Finally, the entropy can be obtained integrating the
thermodynamical relation $dS=T^\mo dM$ and results $S\propto\log M$.

\subsection{$h=1$}
This case has also been considered in \cite{MS}.
The spacetime has constant curvature $-\l^2$, except at the origin.
The \fe define $\y$ up to a constant $\y_0$, thus the most suitable
ansatz for the dilaton and the metric is
\be
\y=\y_0(\a|x|+1),\qquad g={\l^2x^2\over2}+\b|x|+\g.
\ee
Substituting in (\ref{fe1})-(\ref{fe3}), one obtains
\bd
2\b=-2\km,\qquad2\a\g=-\km,\qquad\a\b=\l^2,
\ed
with solution
\bd
\b=-\km,\quad\a=-{\l^2\over\km},\quad\g={(\km)^2\over2\l^2}.
\ed
Hence, after a redefinition of $\y_0$,
\be\lb{h1}
\y=\y_0\left(\l|x|-{\km\over\l}\right),\qquad g={\l^2x^2\over2}-
\km|x|+{(\km)^2\over2\l^2}.
\ee
Notice that the dilaton field equations determine the value of
$\g$ in terms of the other parameters. This fact was not noticed in
\cite{MS}, since the dilaton equations were disregarded there.

Regular black hole solutions exist if $m>0$, with horizons located at
$x_0=\pm\k^2m/\l^2$. In this case the metric has a double zero at $x_0$
and hence a degenerate horizon is present.

The thermodynamical parameters can be computed as in the previous
case and are
\bd
M=0,\qquad T=0.
\ed

\subsection{$h>1$}
The case $h=1$ was somehow degenerate, implying vanishing ADM mass
and temperature. We pass now to consider the
case of $h$ a generic integer. Substituting the ansatz
\be
\y=\l|x|+\b,\qquad g={(\l|x|+\b)^{h+1}+\g\over h+1}\,,
\ee
in the \fe, we obtain
\bd
\b^h=\left(-{\km\over\l}\right),\qquad\g={h-1\over2}\,\b^{h+1}.
\ed
Thus the metric function takes the form
\be
g={1\over h+1}\left[(\l|x|+\b)^{h+1}+{h-1\over2}\,\b^{h+1}\right].
\ee
For $h>1$ odd, the metric is positive definite, and no horizon is
present. For $h>1$ even, horizons are present if $\l>0$, $m<0$,
with $\b=-|\km/\l|^{1/h}$.
The horizons are located at
$$\l x_0=\pm\left|{\km\over\l}\right|^{1/h}\left[\left({h-1\over2}
\right)^{1/(h+1)}+1\right].$$
In this case, the thermodynamical parameters of the \bh are
\bd
M={(h-1)\over4(h+1)}\,\l^{-1/h}(\k^2|m|)^{(h+1)/h},
\ed
\bd
T\propto{\k^2|m|\over 4\p}\propto M^{h/(h+1)},
\qquad\qquad S\propto M^{1/(h+1)}.
\ed
Notice that the ADM mass $M$ is proportional to a power of $m$, and not
to $m$ as one could have expected. The reason is the energy due to the
dilaton coupling.

The previous solutions can be easily generalized to the
case of noninteger $h$, but we shall not discuss in detail this topic.

\section{Multiple sources}
It is interesting to consider the case in which more than one mass
is present. Of course, the solutions are always segments of straight
lines or arcs of convex parabola, which are connected at the location
of the sources with discontinuous derivative. Therefore naked singularities
are avoided only if all the point sources lie in the region included
between two horizons.

We assume that two point sources of mass $m_1$ and $m_2$ are
placed at the points $x_1$ and $x_2$ respectively, and hence
$L_M=-m_1\d(x-x_1)-m_2\d(x-x_2)$, $\ T_{00}=g^\mo[m_1\d(x-x_1)
+m_2\d(x-x_2)]$, $\ T_{11}=0$.

\subsection{$h=0$}
Imposing the ansatz
\bea&\y=\a_1|x-x_1|+\a_2|x-x_2|+\y_0,\cr
&g=\b_1|x-x_1|+\b_2|x-x_2|+\g,
\eea
\begin{figure}
\epsfxsize=0.7\hsize \epsfysize=0.4\hsize
\centerline{\epsfbox{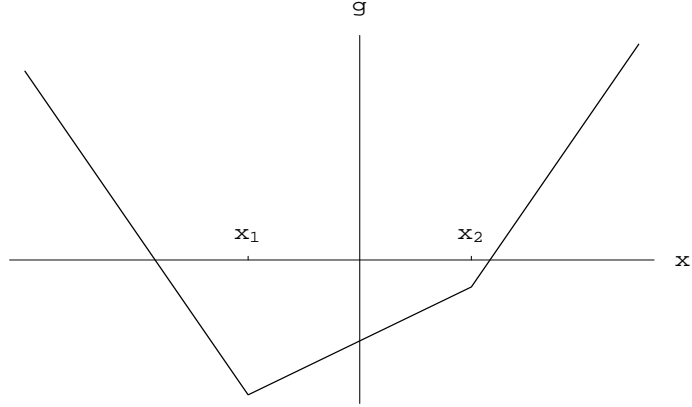}} \caption{\small A typical metric
function $g$ with two sources and two horizons in the case $h=0$.}
\end{figure}
we obtain the conditions
\bd
2\b_1=-2\km_1,\qquad2\b_2=-2\km_2,
\ed
\bd
\k^2\D(m_1\a_2-2m_2\a_1)+2\g\a_1=-\k^2m_1\y_0,
\ed
\bd
\k^2\D(m_2\a_1-2m_1\a_2)+2\g\a_2=-\k^2m_2\y_0,
\ed
\bd
(\a_1+\a_2)(\b_1+\b_2)=\l^2,\qquad(\a_1-\a_2)(\b_1-\b_2)=\l^2,
\ed
where $\D=|x_1-x_2|$. Solutions exist for $m_1\ne m_2$:
\bd
\b_1=-\km_1,\quad\b_2=-\km_2,\quad \a_1=-{\l^2\over\k^2}\,{m_1\over
m_1^2-m_2^2}, \quad\a_2={\l^2\over\k^2}\,{m_2\over m_1^2-m_2^2},
\ed
\bd
\g={3\k^2\D(\mm)\over4},\qquad\y_0={3\l^2\over2\k^2}\,{\D\over\mm}.
\ed
Hence, \be\lb{mbh0}
g=-\km_1|x-x_1|-\km_2|x-x_2|+{3\k^2(\mm)|\xx|\over4}.
\ee
Notice that now, contrary to the case of a single source, $\y_0$ is
determined by the field equations, and $\g$ takes the same value
as in the single particle case for such $\y_0$.

In order to have asymptotically positive metric, we must take negative
values of the masses, as in the single source case.
The solution (\ref{mbh0}) possesses either two or no horizon,
depending on the value of $m_1/m_2$. In particular, for $|m_1|/3<|m_2|
<3|m_1|$, both sources are shielded by a horizon.

\subsection{$h=1$}
Consider now the case $h=1$. The appropriate ansatz is
\bea
&&\y=\y_0\,(\a_1|x-x_1|+\a_2|x-x_2|+1),\cr
&&g={\l^2x^2\over2}+\b_1|x-x_1|+\b_2|x-x_2|+\g.
\eea
We have imposed the vanishing of a term linear in $x$ in the metric.
This is a choice of gauge and is equivalent to fix the position of the
\com of the sources. For definiteness, we assume $x_2>x_1$.

As usual, eq.\ (\ref{fe1}) implies $\b_1=-\km_1$, $\b_2=-\km_2$.
Substituting in (\ref{fe3}), one obtains three independent
equations, that can be cast in the form
\bea\lb{feh1}
&&x_1\a_1+x_2\a_2=0,\cr
&&\k^2(m_1\a_2+m_2\a_1)=-\l^2x_1\a_1,\cr
&&\k^2(m_1\a_1+m_2\a_2)=\l^2(x_1\a_1-1),
\eea
where the first equation is a consequence of our choice of gauge.
From (\ref{feh1}) one obtains
\bd
\a_1=-{\l^2x_2\over\k^2\D(m_1+m_2)},\qquad
\a_2={\l^2x_1\over\k^2\D(m_1+m_2)},
\ed
where $\D=x_2-x_1$.

Moreover, eq.\ (\ref{fe2}) gives rise to two conditions:
\bea
&&(\l^2x_1^2-2\k^2\D m_2+2\g)\a_1+\k^2\D m_1\a_2+\km_1=0,\cr
&&(\l^2x_2^2-2\k^2\D m_1+2\g)\a_2+\k^2\D m_2\a_1+\km_2=0,
\eea
from which one can obtain $\g$ and a further
relation between $x_1$, $x_2$ and the other parameters.
This condition fixes the distance between $x_1$ and $x_2$.
The final result is
\bd
x_1=-{\k^2\m_2^2(\m_1+\m_2)\over\l^2},
\qquad x_2={\k^2\m_1^2(\m_1+\m_2)\over\l^2},
\ed
where $\m_{1,2}=(m_{1,2})^{1/3}$.
One can now write the parameters of the solution in terms of the
mass of the sources only:
\bd
\g={\k^4(\mm)^2\over2\l^2},
\ed
\bd
\a_1={-\l^2\m_1^2\over\k^2(\m_1^5+\m_1^3\m_2^2+\m_1^2\m_2^3+\m_2^5)},
\qquad
\a_2={-\l^2\m_2^2\over\k^2(\m_1^5+\m_1^3\m_2^2+\m_1^2\m_2^3+\m_2^5)}.
\ed
In particular, the metric will take the simple form
\be
{\l^2x^2\over2}-\km_1|x-x_1|-\km_2|x-x_2|+
{\k^4(m_1+m_2)^2\over2\l^2}.
\ee

The only solutions in which both sources are shielded by horizons are
those with $m_1=m_2>0$: in this case the horizons coincide with the location
of the sources.

\begin{figure}
\epsfxsize=0.7\hsize \epsfysize=0.4\hsize
\centerline{\epsfbox{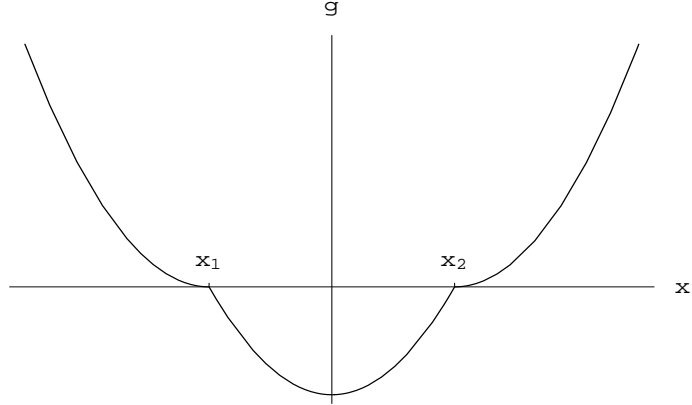}} \caption{\small The metric function
$g$ with $m_1=m_2$ in the case $h=1$.}
\end{figure}

\section{Geodesics}
If one assumes that test particles experience the same conformal
coupling as matter sources, they will not follow the geodesics of
the metric $g_\mn$, but rather those of the rescaled metric
$\hat g_\mn=\y^2 g_\mn$.
This is obvious if one writes the matter action in the form,
equivalent to (\ref{LM})
\be
\int d^2x\sqrt{-g}\,\y\,L_M=-m\int\y\,ds=
-m\int\sqrt{\y^2\,g_\mn\dot x^\m\dot x^\n}\,ds,
\ee
where the dot indicates a derivative \wr the proper time $s$.

Varying this action, the geodesic equations can also be written as
\bd
{D^2 x^\m\over ds^2}+{\de\ln\y\over\de x^\n}
\left(2{Dx^\m\over ds}{Dx^\n\over ds}-g^\mn\right)=0
\ed
where $D/ds$ denotes the covariant derivative evaluated in the metric
$g_\mn$.

In terms of the metric function $g(x)$, the previous equations become
\bea
&&\ddot x-{g'\over2g}\,\dot x^2-\ha gg'\dot t^2-{\y'\over\y}(g-\dot
x^2),\cr
&&\ddot t+\left({g'\over g}+2{\y'\over\y}\right)\dot x\dot t=0.\nn
\eea
These equations can be easily integrated, giving
\be\lb{tdot}
\dot t={E\over \y^2\,g},\qquad\qquad\dot x={1\over \y^2}\sqrt{E^2+
\e\y^2g},
\ee
with $\e=0$ for massless particles, and $\e=1$ for massive ones, while
$E$ is the energy of the test particle.

One can perform a further integration of (\ref{tdot}).
After simple manipulations one obtains the relations
\be\lb{int}
s=\int{\y^2\over\sqrt{E^2+\e\y^2g}}\ dx,\qquad
t=\int{E\over g\sqrt{E^2+\e\y^2g}}\ dx,
\ee
that permit to compute $x$ as a function of $s$ and $t$,
respectively.

These integrals can be easily evaluated for massless particles.
For the solution (\ref{h0}), corresponding to $h=0$, one gets,
for positive $x$,
\bd
x=\left(3\k^4m^2E\over\l^4\right)^{1/3}s^{1/3}+{\km\over\l^2}\y_0
={\km\over2\l^2}\y_0\left(1-e^{-\km t}\right).
\ed
It follows that the
horizon is reached in a finite  proper time $s$, but in an
infinite coordinate time $t$. The integrals (\ref{int}) cannot be
evaluated analytically for massive particles, but it is easy to
check that the qualitative behavior of the geodesics is the same
as for massless particles.

For $h=1$, the solution (\ref{h1}) yields for massless particles,
$x>0$,
\bd
x=\left(3E\over\l^2\y_0^2\right)^{1/3}s^{1/3}+{\km\over\l^2}=
-{2\over\l^2t}+{\km\over\l^2}.
\ed
Again, the integrals cannot be
explicitly evaluated for massive particles. However, it is easily
seen that the behavior of the geodesics is essentially the same as
in the previous case.

To compute the force experienced in the Newtonian limit by a test
particle in the field generated by a point particle, we assume
that the relevant spatial coordinate is that in which the metric
$\hat g_\mn$ takes the Schwarzschild form
$ds^2=-A\ dt^2+A^\mo dr^2$. This is obtained by defining a new
coordinate $r=\int\y^2dx$.
For $h=0$, for example, from (\ref{h0}), with $m<0$,
\bd r={\k^2|m|\over3\l^2}\y^3,\ed
and
\bd
A=3\k^2|m|\left[r-\left(9\k^2|m|\over\l^2\right)^{1/3}\y_0\,r^{2/3}\right].
\ed
The gravitational potential displays a term linear in
$r$ and a correction proportional to $\y_0$. Deriving, one obtains
that the force has a constant component proportional to the mass,
with a short-range correction diverging at the location of the source.

\section{Conclusions}
We have studied the complete static solutions of two-dimensional
dilaton-gravity theories in the presence of single or multiple
sources. The dilaton equations constrain the parameters of the
solutions, so that no regular \bh solutions exist for odd $h$,
except the case $h=1$, where they assume a degenerate form.

We also have discussed the action of gravity on test particles.
Since these should be conformally coupled, the force exerted by
a point mass is not constant, as one would naively expect, except
at large distances.

\end{document}